\documentclass[english,keywords,aps,amsmath,amssymb,twocolumn]{revtex4-1}
\usepackage[T1]{fontenc}
\usepackage[latin1]{inputenc}
\usepackage{babel}
\usepackage{amssymb}
\usepackage{graphicx}
\usepackage{color}
\usepackage{bm}
\usepackage{longtable}
\usepackage{amsmath} 
\usepackage{amsfonts}
\usepackage{amssymb}
\begin{document}
\title{ Probing various formulations  of macrorealism for unsharp quantum measurements}
\author{Swati Kumari }
\author{A. K. Pan \footnote{akp@nitp.ac.in}}
\affiliation{National Institute of Technology Patna, Ashok Rajpath, Patna, Bihar 800005, India}
\begin{abstract}
Standard Leggett and Garg inequalities (SLGIs) were formulated for testing incompatibility between the classical world view of macrorealism and quantum mechanics. In recent times, various other formulations, such as, Wigner form of LGIs (WLGIs), entropic LGIs (ELGIs) and the no-signaling in time (NSIT) condition have also been proposed. It is also recently argued that \textit{no} set of SLGIs  can provide the necessary and sufficient conditions for macrorealism but a suitable conjunction of NSIT conditions provides the same. In this paper, we first provide a comparative study of the various formulations of LGIs for testing macrorealism pertaining to the two different unsharp measurements. While the violations of WLGIs are more robust than SLGIs and ELGIs for spin-POVMs, here we demonstrate that for the case of biased POVMs, the quantum violations of both SLGIs and ELGIs provide the same robustness as WLGIs. Importantly, the violations of all formulations of LGIs can be achieved for \textit{any non-zero value} of unsharpness parameter. We have also studied the connection between LGIs and NSIT conditions.  Further, we investigate  the role of the  joint measurability of the POVMs in the violation of LGIs and found that there is \emph{no} generic connection.
\end{abstract}
\maketitle
\section{Introduction}
Even almost eighty years after the Schroedinger famous cat paradox experiment, it is still a debatable issue, how the realist view of macroscopic classical world emerges from the framework of quantum mechanics(QM). The macrorealist view asserts that the properties of objects exist at all instant of time, and are independent of the observation. In this regard, the central question is whether such a macroscopic world view is compatible with the statistics of QM. In 1985, Leggett and Garg \cite{leggett85} formulated an inequality which is assumed to be obeyed by a macrorealist theory, provides an elegant scheme for experimentally testing the compatibility between the classical world view of macrorealism and QM. 

The notion of macrorealism consists of two main assumptions \cite{leggett85,leggett,A.leggett} which are in principle valid in our everyday world are the following;
	
	\emph{ Macrorealism per se (MRps):} If a macroscopic system has two or more macroscopically distinguishable ontic states available to it, then the system remains in one of those states at all instant of time.
	
	\emph{Non-invasive measurability (NIM):} The definite ontic state of the macrosystem is determined without affecting the state itself or its possible subsequent dynamics.\\
	
	Based on the above two assumptions, the standard Leggett and Garg inequalities (SLGIs) was derived. Such inequalities can be  violated in certain circumstances, which thereby imply that one or both the assumptions of MRps and NIM are not compatible with all the quantum statistics. Since then based on various theoretical proposals \cite{budroni, budroni15,maroney,kofler,clemente,UshaDevi,saha,moreira,clemente16,hall,mal} quite a number of experiments \cite{a.p,goggin,xu,dressel,suzuki,arndt,gerlich,julsgaard,isart,souza,mahesh11,katiyar13,formaggio,katiyar} have been performed. Leggett and Garg initially proposed an rf-SQUID flux qubit as a promising system to test their inequalities \cite{leggett85}. Palacious-Layloy et al.\cite{a.p} performed an experiment using the superconducting qubit with continuous weak measurement which confirmed the violation of a LGI. Their experiment \cite{a.p} was followed by a  number of LGI tests using different physical systems such as photons \cite{goggin,xu,dressel,suzuki}, heavy molecules\cite{arndt,gerlich} and quantum optical systems in combination with atomic gases\cite{julsgaard} or massive objects \cite{isart}. Recently the violations of SLGI is experimentally shown for neutrino oscillations \cite{formaggio} and for a $3$-level system \cite{katiyar}.
	
Besides SLGIs, there have been other interesting  formulations for testing the macrorealism, such as, Wigner form of Leggett-Garg inequalities (WLGIs)\cite {saha}, entropic formulation of Leggett-Garg inequalities (ELGIs) \cite{UshaDevi} and no-signaling in time (NSIT) \cite{clemente,clemente16}.  The NSIT condition is considered to be the necessary condition for macrorealism\cite{kofler} and seems to be analogus to the no-signaling condition in Bell's theorem.  Although Bell's inequalities are structurally analogous to SLGI, but it is recently shown by Budroni and Emary\cite{budroni} that the SLGI can even be violated upto its algebraic maximum within the framework of QM. Such amount of violation of Bell's inequalities can only be achieved for post-quantum theories. In an interesting paper, Clemente and Kofler \cite{clemente16} have argued that no set of LGIs can provide necessary and sufficient condition for macrorealism in contrast to the case of CHSH inequalities providing the same for local realism\cite{fine}. A suitable conjunction of two-time and three-time NSIT conditions provide necessary and sufficient condition for macrorealism \cite{clemente16}. By noting this fact, it is claimed \cite{clemente} that NSIT is the better candidate for testing macrorealism than LGIs.

In this paper, we first study the violation of various formulations of LGIs for the case of unsharp measurements. In particular, we compared the quantum violations of SLGIs, WLGIs and ELGIs for two different unsharp measurements when the measurements are performed at three different times. It is recently argued \cite{saha} that the violations of WLGIs are more robust than  SLGIs for spin-POVMs. This is due to the fact that the former can be violated for lower values of sharpness parameter than the  later. By considering the biased POVMs, we demonstrate here that the SLGIs and ELGIs provide the same robustness as WLGIs. Importantly, the violations of all the three types of LGIs can be achieved for \textit{any non-zero value} of the unsharpness parameter. Thus, if LGI is considered to be an indicator of classicality of macroscopic system them any arbitrary unsharp measurement does not lead classicality through LGI. 

Further, we have re-examined the relation between LGIs, NSIT conditions and macrorealism for the case of sharp measurement. It is already pointed out\cite{clemente} that even if \emph{all} the NSIT conditions are violated, the SLGI  may not be violated. But, if  SLGI is violated then at least one of the NSIT conditions is required to be violated. Similar to SLGIs, the three-time NSIT conditions can be shown to be necessary for WLGIs but not for macrorealism.  We have shown that pertaining to the three-time LG scenario considered here, if at least one of the  NSIT conditions is violated then one of the $24$ WLGIs will also be violated except for the instants when $\tau=\pi/4$, for  $\rho=|+\rangle\langle+|$ or $\rho=I/2$. We provide an explanation why NSIT conditions are better criteria than LGIs. This is done by invoking the notion of disturbance caused to the subsequent measurement due to a prior measurement.

We have also investigated the possible connection between the joint measurability and the violation of LGIs, similar to the connection between the local  joint measurability and CHSH inequality \cite{fine}. We show that there is no generic connection between the violation of LGIs and joint measurability.

This paper is organized as follows. In Section II, we provide a comparative study of violations of various formulations of Leggett-Garg inequalities, viz., SLGIs, WLGIs and ELGIs for two unsharp measurements, and demonstrate that unsharp measurement does not lead to classicality in general. We then examine the relation between the NSIT conditions and the violation of LGIs in Section III.  In Section IV, we probe the possible  connection between  joint measurability and the violation of LGIs.  We summarize and discuss our results in Section V.
	
\section{Violation of various Leggett-Garg inequalities for unsharp measurements}
Let at time $t_1$, an ensemble of similarly prepared macroscopic system has two ontic states available to it  and evolves from one state to another with time. However, at any particular instant the system is found to be in a definite macroscopic state. Now, at all instant of time the measurement of a suitable dichotomic observable $\hat{M}$ should produce definite outcomes $+1$ or $-1$ according to the assumption of MRps . Let the measurement of $\hat{M}$ is performed on the macroscopic system at three different times $t_1$, $t_2$ and $t_3 (t_3>t_2>t_1)$ which in turn implying that measurement observables $\hat{M_1}$, $\hat{M_2}$ and $\hat{M_3}$ respectively in the Heisenberg Picture.

Now, the notion of NIM assumes that the measurement of $\hat{M_1}$ can in principle be non-invasive, so that, the measurement of $\hat{M_2}$ at $t_2$ or $\hat{M_3}$ at $t_3$ remains unaffected due to the measurement of $\hat{M_1}$ and similarly for the other set of  sequential measurements. In other words, the NIM implies that the existence of  joint probabilities of different outcomes $P(M_1^{\pm},M_2^{\pm},M_3^{\pm})$ and the relevant marginals are unaffected by the prior or future measurements.

 By using the MRps and NIM assumptions, the following inequality can be derived,
\begin{eqnarray}
\label{lgi}
\Delta_{s}^{LG}&=&\langle M_1 M_2\rangle+\langle M_2 M_3\rangle-\langle M_1 M_3\rangle\leq1
\end{eqnarray}
which is the well-known standard LGI \cite{leggett85,leggett,A.leggett}, obeyed by a macrorealist theory. By relabeling the measurement outcomes of each $M_{i}$ as $M_{i}=-M_{i}$ with $i=1,2$ and $3$, three more standard LGIs can be obtained.  In order to examine the  empirical validity of ineq.(\ref{lgi}) in the framework of QM, let us consider a state in two-level system $\rho(t_1)=|\psi(t_1) \rangle \langle \psi(t_1)|$ at $t_1$, where
\begin{equation}
\label{state}
|\psi(t_1)\rangle = \cos\theta |0 \rangle + e^{i\phi} \sin\theta |1 \rangle
\end{equation}
with $\theta \in [0,\pi]$, $\phi \in [0,2\pi]$ and the measurements of unsharp observables $\hat{M_1}$, $\hat{M_2}$ and $\hat{M_3}$ at three times $t_1$, $t_2$ and $t_3$ respectively.
The system evolves under unitary operator $U_{|{i-j}|\Delta t}$ in the time interval between $t_i$ and $t_j$ where $i,j = 1,2,3$ with $i> j$.\\
\\ 
For our purpose, we consider the sequential measurements of general POVMs is of the form
\begin{equation}
\label{GO}
M_{i}^{\pm}(x,\vec{m_{i}})=\frac{{\mathbb I} \pm (x{\mathbb I}+\vec{m_{i}}.\sigma)}{2}
\end{equation}
with $|x|+|\vec{m_{i}}|\leq1$, where $|x|$ is the biasedness and $|\vec{m_{i}}|$ is the sharpness parameter. Note that, Eq.(\ref{GO}) reduces to the spin-POVMs when $x=0$. 

At time $t_1$, we consider the POVMs as $M_{1}^{\pm}(x,\vec{m_1})$with $\vec{m_1}=\eta \vec{z}$. The time evolution of $M_{1}^{\pm}(x,\vec{m_1})$ in two different times $t_2$ and $t_3$ are given by $M_{2}^{\pm}(x,\vec{m_2})=U_{\Delta t}^{\dagger} M_{1}^{\pm}(x,\vec{m_1}) U_{\Delta t}$  and $M_{3}^{\pm}(x,\vec{m_3})=U_{2\Delta t}^{\dagger} M_{1}^{\pm}(x,\vec{m_1}) U_{2\Delta t}$ respectively. If intermediate unitary evolution is taken to be $U_{|{i-j}|\Delta t}= exp(-i \omega\sigma_{x}|{i-j}|\Delta t)$, then $M_{2}^{\pm}(x,\vec{m_2})$ and $M_{3}^{\pm}(x,\vec{m_3})$ can be written as $M_{2}^{\pm}(x,\vec{m_2})=\frac{\mathbb{I}\pm(x\mathbb{I}+\vec{m_2}.\sigma)}{2}$ and $M_{3}^{\pm}(x,\vec{m_3})=\frac{\mathbb{I}\pm(x\mathbb{I}+\vec{m_3}.\sigma)}{2}$  respectively,
where $\vec{m_2} = \eta(\sin 2\tau \vec{y} + \cos 2\tau \vec{z})$ and $\vec{m_3} = \eta (\sin 4\tau \vec{y }+ \cos 4\tau \vec{z})$.

The probability of an outcome, say $M_1=+1$, is then given by $tr(\rho(t_1)M_1^{+})$, for which the post-measured density matrix can be written as $\rho_{+}(t_{1})=(\sqrt{M_{1}^{+}}\rho_0(t_1)\sqrt{M_{1}^{+}}^{\dag})/tr(\rho_0(t_1)M_{1}^{+})$. Subsequently, the post-measurement state evolves under the unitary operator $U_{\Delta t}= exp(-i \omega\sigma_{x}\Delta t)$  to the state $\rho_{+}(t_2)= U_{\Delta t} \rho_{+}(t_1)U_{\Delta t}^{\dagger}$ at a later instant $t_2$ where $\Delta t = t_2-t_1$. For notational simplicity, we shall use $\Delta t=\tau$.

The joint probability of different outcomes for two POVMs can  then be calculated by using the formula is given by
\begin{eqnarray}
\label{pb}
&&P(M_i^{k}(x,\vec{m_i}),M_j^{l}(x,\vec{m_j}))\\
\nonumber
&=&Tr(U_{|i-j|\Delta t}\sqrt{M_i^{k}}\rho(t_i)\sqrt{M_i^{k}}^\dag U_{|i-j|\Delta t}^\dag M_j^{l})
 \end{eqnarray}
where $k,l=\pm$. 
Henceforth, for avoiding the clumsiness of the notation, we denote $M_{i}^{l}(x,m_i)$ as $M_{i}^{l}$ and so on. We now proceed to study the various formulations of LGIs for unsharp measurements described by the POVMs given by Eq.(\ref{GO}). 
\subsection{Violation of SLGI for unsharp measurements}
 In order to examine the compatibility between SLGI and QM we calculate the quantum mechanical value($\Delta_{s}^{Q}$) of the LHS of ineq.(\ref{lgi}) for the state $|\psi(t_1)\rangle$ given by Eq.(\ref{state}). Joint expectation value $\langle M_i M_j \rangle$ can be obtained by calculating the joint probabilities given by Eq.(\ref{pb}) for the state $|\psi(t_1)\rangle$, so that,  $\langle M_i M_j \rangle= \sum_{k,l=\pm} k l P(M_i^{k},M_j^{l})$. For the POVMs given by Eq.(\ref{GO}), $\Delta_{s}^{Q}$ corresponding to ineq.(\ref{lgi}) can be obtained as
\begin{widetext}
\begin{eqnarray}
\Delta_{s}^{Q}&=&\frac{1}{8} [8 x^2+\eta( (4 \sin 2 \theta  \sin 2 \tau \sin \phi((4 x \cos ^2\tau) \\
\nonumber
&+&(2 \cos 2 \tau -1) (\sqrt{(1-x)^2-\eta^2}-\sqrt{(1+x)^2-\eta^2}))-8 \eta (\cos 4 \tau -2 \cos 2 \tau ))\\
\nonumber
&+&2\sin\phi\cos2\theta\cos4\tau(\sqrt{(1-x)^2-\eta^2}-\sqrt{(1+x)^2-\eta^2}) \\
\nonumber
&-&\sin2\theta\sin4\tau(\sqrt{(1-x)^2-\eta^2}-\sqrt{(1+x)^2-\eta^2}))\\
\nonumber
&+&2 \eta \cos 2 \theta (4 x \sin ^2 2 \tau +8 x \cos 2 \tau +(\sqrt{(1-x)^2-\eta^2}-\sqrt{(1+x)^2-\eta^2}))]
\end{eqnarray}
\end{widetext}
Now, for $x=0$, we obtain the results \cite{saha} of unbiased spin-POVMs $M_{1}^{\pm}(0,\eta)=(\mathbb{I}\pm \eta\sigma_{z})/2$, is given by 
\begin{eqnarray}
\label{sl1}
\Delta_{s}^{Q}&=&\eta ^2 \left(2 \cos 2 \tau -\cos 4 \tau \right)
\end{eqnarray}
The violation of SLGI  can be obtained  upto the values of $\eta>0.81$ for a range of $\tau$ and the maximum violation is obtained for sharp measurement ($\eta=1$) at $\tau=\pi/3$. Note that, similar to the case of sharp measurement, the expression of the quantity $\Delta_{s}^{Q}$ in Eq.(\ref{sl1}) is also state independent.

Next, for another choice of $x=\eta-1$, the POVMs takes the form $M((\eta-1),\eta\vec{z})=\eta(I+\sigma_{z})/2$ \cite{quintino}. The expression of $\Delta_{s}^{Q}$ is given by
\begin{widetext}
\begin{eqnarray}
\label{sl2}
\Delta_{s}^{Q}&=&\frac{1}{8} [\eta  (4 \sin 2 \theta  \sin 2\tau \sin\phi ) (4 (\eta -1) \cos ^2\tau + 2 \sqrt{1- \eta }(2 \cos 2 \tau -1))\\
\nonumber
&-& 4\eta\sqrt{1-\eta } (\sin \phi \sin 2 \theta \sin4 \tau +\cos2 \theta \cos 4\tau)+2 \eta  \cos 2 \theta  (4 (\eta -1) \sin ^2 2\tau \\
\nonumber
&+&8 (\eta -1) \cos 2 \tau + 2\sqrt{1-\eta })+8 (\eta -1)^2-8\eta^2(-2 cos 2\tau +cos 4\tau)]
\end{eqnarray}
\end{widetext}
which explicitly depends on the parameter $\theta$ and $\phi$ of the state $|\psi(t_1)\rangle$. Curiously, for  the values of $\theta=\pi/3$, $\phi=\pi/2$ and $\tau=5\pi/6$, the Eq.(\ref{sl2}) reduces to the simple form $\Delta_{s}^{Q}=1+\eta^2/2$. Thus, the violation of SLGI can be obtained for \textit{any non-zero value} of $\eta$. This feature is in contrast to the case of spin-POVMs where the violation is obtained only when $\eta>0.81$. Hence, we can conclude that the degree of unsharpness of the measurement  does \textit{not} play an important role for the violation of LGIs for the qubit system. 

\subsection{Violation of WLGIs for unsharp measurement}
We now study the violation of Wigner form of Leggett-Garg inequalities (WLGIs), which is recently introduced by Saha \textit{et al.}\cite{saha}. Wigner form of local realist inequality \cite{ep}  is derived based on the locality condition and the existence of the joint probability distributions for the occurrence of different possible combinations of the outcomes of measurements of the relevant observables.  Using the NIM condition that the overall joint probabilities and their marginals would remain unaffected by the measurements, the WLGI can be derived as follows. For example, the joint probability $P(M_2^{+},M_3^{-})$ of obtaining the outcomes for the sequential measurements at two instants $t_2$ and $t_3$ can be obtained by marginalization of $\hat{M_1}$ is given by
\begin{eqnarray}
P(M_2^{+},M_3^{-})&=&\sum_{i=\pm}P(M_1^{i},M_2^{+},M_3^{-})
\end{eqnarray}
Writing similar other expressions for the joint probabilities $P(M_1^{+},M_2^{+})$ and $P(M_1^{-},M_3^{-})$, we get
$P(M_1^{+},M_2^{+})+P(M_1^{-},M_3^{-})-P(M_2^{+},M_3^{-})=P(M_1^{+},M_2^{+},M_3^{+})+P(M_1^{-},M_2^{-},M_3^{-})$.
Invoking the non-negativity of the probability, the following form of inequality is obtained in terms of three pairs of two-time joint probabilities, is given by 
\begin{eqnarray}
\label{wlgi}
\Delta_{w}^{LG}&=&P(M_{2}^{+},M_{3}^{-})-P(M_{1}^{+},M_{2}^{+})\\
\nonumber
&-&P(M_{1}^{-},M_{3}^{-})\leq 0		
\end{eqnarray}
which is termed as WLGI. Note that, $23$ more such inequalities can also be derived in this manner \cite{saha}.

In order to showing the quantum violation of ineq.(\ref{wlgi}) we calculate the quantum mechanical expression ($\Delta_{w}^{Q}$) of the LHS of ineq.(\ref{wlgi}). The expression of  $\Delta_{w}^{Q}$ is given by
 \begin{widetext}
\begin{eqnarray}
\Delta _{w}^{Q}&=&\frac{1}{16}[ (-4 \left(3 x^2+1\right)+\eta[(-4 \sin2\theta\sin2\tau\sin\phi((x+1) \cos 2 \tau + \sqrt{(1+x)^2-\eta^2} +x-1)\\
\nonumber
&+&2\cos 2\theta \cos 4 \tau \sin \phi(\sqrt{(1+x)^2-\eta^2} -2 \sqrt{(1+x)^2-\eta^2})\\
\nonumber
&+&2\sin 2\theta\sin 4\tau \sqrt{(1+x)^2-\eta^2}-4\eta(2 \cos 2 \tau +\cos 4 \tau )]\\
\nonumber
&+&2 \eta \cos 2 \theta (\cos 4 \tau -x (4 \cos 2 \tau +3 \cos 4 \tau +5)+\sqrt{(1+x)^2-\eta^2} -1))]
\end{eqnarray}
\end{widetext}
For spin-POVMs (for $x=0$), the  above expression of $\Delta_{w}^{Q}$ can be written as \cite{saha}
\begin{eqnarray}
\label{wl1}
\nonumber
\Delta_{w}^{Q} &=&\frac{1}{16} [\eta(-4 \sin 2 \theta  \sin 2 \tau  \sin \phi  (\sqrt{1-\eta^2 } +\cos 2 \tau -1)\\
&-&2\sqrt{1-\eta^2 }(\cos 2\theta\cos 4\tau-\sin\phi\sin 2\theta\sin 4\tau)\\
\nonumber 
&-&4 \eta  (2 \cos 2 \tau +\cos 4 \tau ))\\
\nonumber
&+&2\eta\cos 2 \theta(\sqrt{1-\eta^2 }+\cos 4 \tau -1)-4]
\end{eqnarray}
In contrast to the $\Delta_{s}^{Q}$ given by the Eq.(\ref{sl1}), $\Delta_{w}^{Q}$ is dependent on state. The ineq.(\ref{wlgi}) is violated for a ranges of values of $\theta$, $\phi$, $\tau$ and $\eta$. The lowest value of $\eta$ is possible at $\theta=\pi/3$, $\phi=\pi/2$ and $\tau=\pi/3$, for which $\Delta_{w}^{Q}=(3\eta(1+\eta-\sqrt( 1-\eta^2))-2)/8$. It is seen that the  violation of ineq.(\ref{wlgi}) is obtained  for the values $\eta > 0.69$. We have checked that none of the $24$ WLGIs is violated for spin-POVMs if $\eta \leq 0.69$.
 Note here that, the violation of SLGI given by ineq.(\ref{lgi}) was obtained for spin-POVMs for the value of the sharpness parameter $\eta> 0.81$.  Then WLGIs can be violated between the ranges of $0.69-0.81$ of the sharpness parameter where SLGI is not violated. By noting this feature it is argued \cite{saha} that the violation of WLGI can be considered to be more robust than the violation of SLGI. We examine here that if the conclusion remains same for other form of unsharp measurement.

For this, we consider another form of unsharp measurement (biased POVMs) by taking $x =\eta-1$ in Eq.(\ref{GO}) and choose a suitable inequality form $24$ WLGIs is given by 
 \begin{eqnarray}
\label{wlgi1}
\Delta_{w}^{LG}&=&P(M_{1}^{+},M_{3}^{-})-P(M_{1}^{+},M_{2}^{-})\\
\nonumber
&-&P(M_{2}^{+},M_{3}^{-})\leq 0		
\end{eqnarray}
 The QM expression $\Delta_{w}^{Q}$ of LHS of ineq.(\ref{wlgi1}) for biased POVMs is given by
\begin{eqnarray}
\label{wlgi2}
\Delta_{w}^{Q}&=&\frac{\eta}{8} \Big[2 (\sin 2 \theta  \sin 2 \tau \sin\phi  (\eta \cos 2\tau +\eta -2)\\
\nonumber
&+&2 \eta  \cos 2\tau -\eta  \cos 4\tau +\eta -2)\\
\nonumber
&+&\cos 2\theta (2 \eta  \sin ^2(2 \tau )+4 (\eta -1) \cos 2 \tau )\Big]
\end{eqnarray}
 For the values of $\theta=\pi/3$, $\phi=\pi/3$ and $\tau=5\pi/6$, the  Eq.(\ref{wlgi2}) takes  simple form $\Delta_{w}^{Q}=\eta^2/8$. Then,  WLGI can also be violated for \textit{any non-zero value of \textit{sharpness parameter}}($\eta$). Hence, the violation of both WLGIs and SLGIs provide the \textit{same} robustness for the biased POVMs considered here. 
\subsection{Entropic Leggett-Garg Inequality for three observables}
We now probe the violation of the entropic formulation of Leggett-Garg inequality (ELGI). Such an inequality can be derived  by using the properties of Shannon entropy from the classical information theory, viz., the chain rule $H(M_{i},M_{j})=H(M_{i}|M_{j}) + H(M_{j}) = H(M_{j}|M_{i}) + H(M_{i})$ and $H(M_{i},M_{j})\leq H(M_{i})+H(M_{j})$. The latter implies
that total information of individual random variables cannot be less than the information carried by joint variables. We then have
$ H(M_{i}|M_{j}) \leq H(M_{i})$, i.e.,
the information possessed by a random variable decreases if a condition is imposed. 

 For our purpose, we consider the joint Shannon entropy for three observables $\hat{M_1}$, $\hat{M_2}$ and $\hat{M_3}$  at three different instants, say, $t_1$, $t_2$ and $t_3$ respectively.  Using the chain rule for joint Shannon entropy one has
\begin{eqnarray}
\label{chrule}
\nonumber
H(M_{1},M_{2},M_{3})&=&H(M_{3}|M_{2},M_{1})+H(M_{2}|M_{1})+H(M_1)
\end{eqnarray}
Using other properties of Shannon entropy, we have $H(M_{1},M_{3})\leq H(M_{1},M_{2},M_{3})$, and $H(M_{1}|M_{2},M_{3})\leq H(M_{1},M_{2})-H(M_{2})$. Writing all the quantities in this manner and rearranging the terms, one obtains ELGI is of the form
\begin{eqnarray}
\label{MRe}
\Delta_{e}^{LG}&=&H(M_{1},M_{3})-H(M_{2},M_{3})\\
\nonumber
&-&H(M_{1},M_{2})+H(M_{2})\leq 0
\end{eqnarray}
where $H(M_{i},M_{j})= -\sum_{k,l=\pm} P(M_{i}^{k},M_{j}^{l})\ln P(M_{i}^{k},M_{j}^{l}) $ where $i,j = 1,2,3$.
and $H(M_{i})= -\sum_{k=\pm} P(M_{i}^{k})\ln P(M_{i}^{k}) $ where $ i = 1,2$. Similarly, two more ELGIs can be derived.

The ELGI given by ineq.(\ref{MRe}) is violated for the spin-POVMs ($x=0$) for the choices of $\phi=\pi/2$, $\theta= 1.7$, for the values of $\eta > 0.972$ (Figure $1$). The violation of ineq.(\ref{MRe}) cannot be obtained for $\eta < 0.972$ for any choice of $\theta$, $\phi$ and $\tau$.
\begin{figure}[h]
{\rotatebox{0}{\resizebox{7.0cm}{5.0cm}{\includegraphics{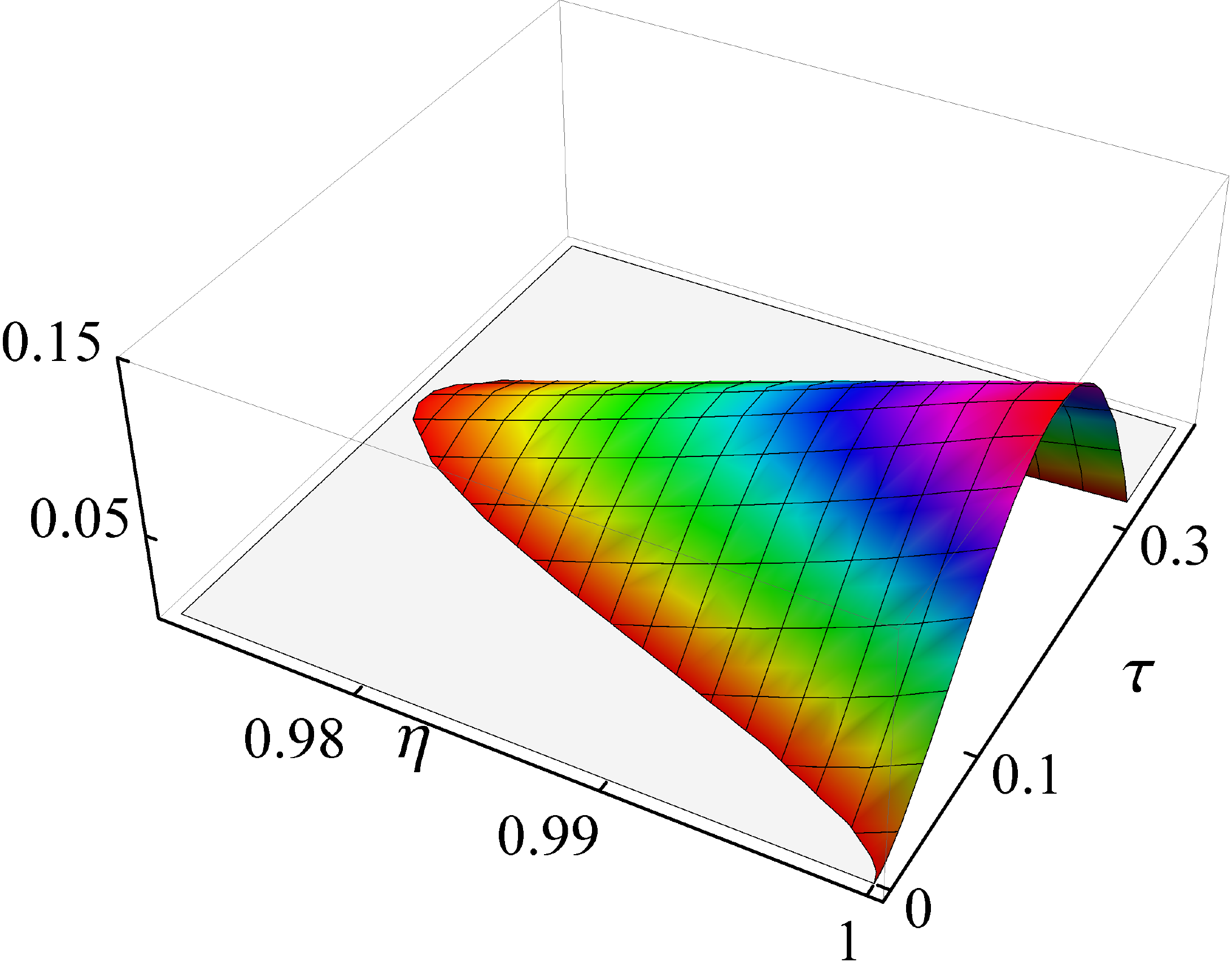}}}}
\caption{Plot for showing  violation of ELGI given by ineq.(\ref{MRe}) for the case of spin-POVMs ($x=0$). The values of the relevant parameters are taken to be $\theta= 1.7$ and $\phi=\pi/2$.} 
\end{figure}

If we consider the POVMs of the form $M((\eta-1),\eta\vec{z})=\eta(I+\sigma_{z})/2$ by putting $x=\eta-1$ in Eq.(\ref{GO}), the ELGI given by ineq.(\ref{MRe}) is violated \textit{for all values of $\eta$} for the choices of values of $\theta=1.7$ and $\phi=\pi/2$ (Figure $2$). Hence, for biased POVMs the violations of both SLGI and ELGI can be achieved for \textit{any non-zero value} of the sharpness parameter($\eta$) thereby providing the same robustness as the violation of WLGI.
\begin{figure}[h]
{\rotatebox{0}{\resizebox{7.0cm}{5.0cm}{\includegraphics{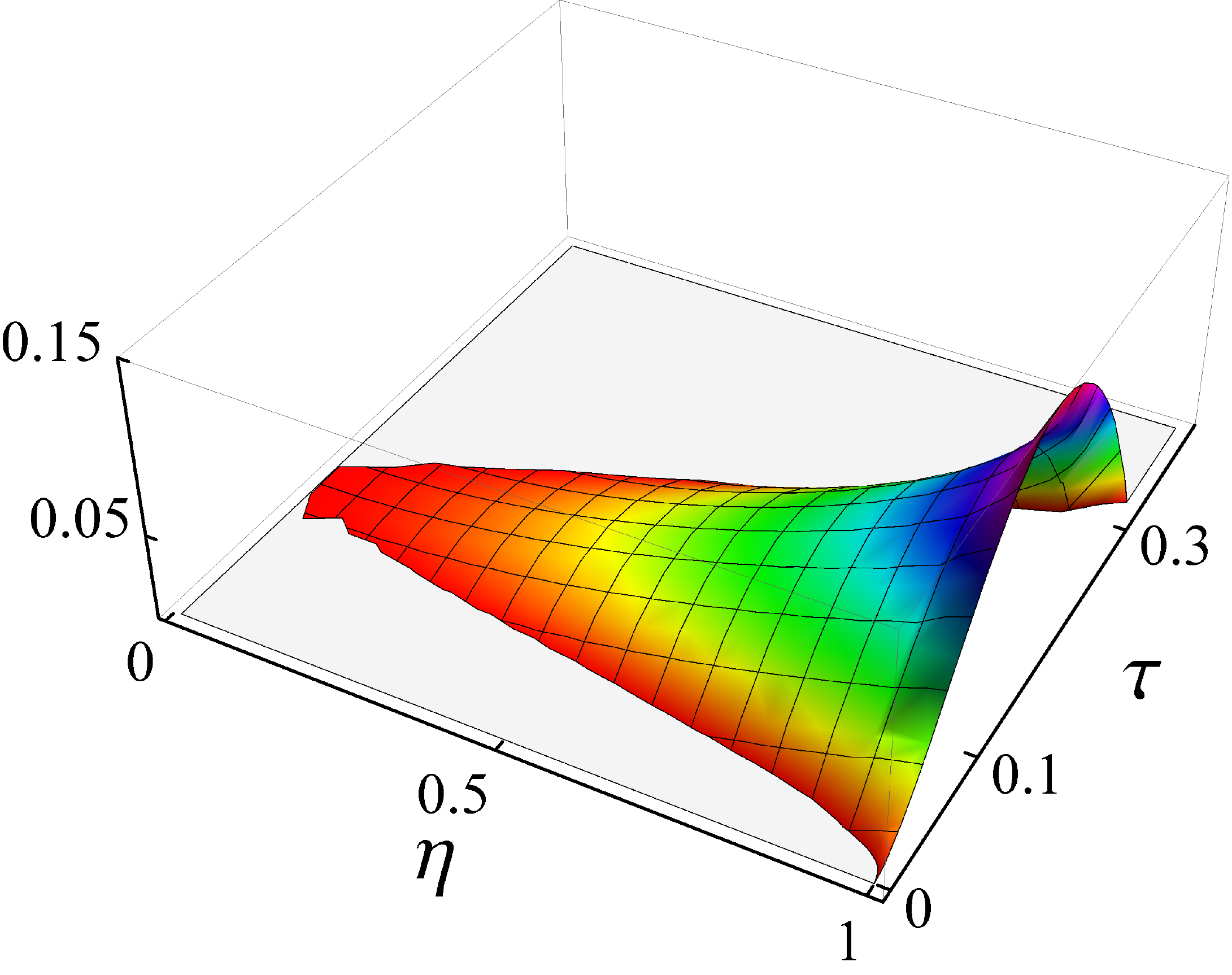}}}}
\caption{Plot for showing the violation of ELGI given by ineq.(\ref{MRe}) for any non-zero value of $\eta$ for the case of biased-POVMs ($x=\eta-1$). The values of the relevant parameters are taken to be $\theta= 1.7$ and $\phi=\pi/2$.} 
\end{figure}

 We can then argue that if the violation of LGI is considered as an indicator of non-classicality, then every unsharp measurement does \textit{not} lead to classicality for qubit system. 

\section{No-signaling in Time, LGIs and Macrorealism}
The no-signaling in time (NSIT) condition assumes that the probability of obtaining an outcome of the measurements remains unaffected due to the prior measurements. It is analogous to the no-signaling in space condition in Bell's theorem and can be considered as the statistical version of NIM condition. Note however that while the violation of no-signaling in space-like separated measurements leads to a contradiction with special theory of relativity, the violation of NSIT does not produce any such inconsistency. The conjunction of all the NSIT conditions ensures that the existence of global joint probability distribution $P(M_{1}, M_{2} , M_{3})$. If NSIT condition is violated at the statistical level, such a violation can be extrapolated at the level of individual measured value, implying that the NIM condition is violated.  In view of Leggett\cite{A.leggett}, the NIM naturally includes MRps condition. However, Clemente and Kofler\cite{clemente} introduces the concepts of strong and weak NIMs and argues that it is the strong NIM which implicitly assumes MRps.
	
A general two-time NSIT condition can be read as 
\begin{equation}
NSIT_{(i)j}:P(M_{j})=\sum_{M_{i}} P(M_{i},M_{j})
\end{equation}
 which means that the probability of obtaining a particular outcome of the measurement of $M_{j}$  is unaffected by the prior measurement  $M_{i}$.\\

 Then the two-time $NSIT_{(1)2}$, $NSIT_{(1)3}$ and $NSIT_{2(3)}$ conditions are respectively given by
\begin{equation} 
NSIT_{(1)2}:P(M_{2})=\sum_{M_{1}} P(M_{1},M_{2})
\end{equation}
\begin{equation}
NSIT_{(1)3}:P(M_{3})=\sum_{M_{1}} P(M_{1},M_{3}).
\end{equation}
\begin{equation}
NSIT_{(2)3}:P(M_{3})=\sum_{M_{2}} P(M_{2},M_{3}).
\end{equation}
Similarly, three-time condition $NSIT_{(1)23}$ states that  the joint probability $P(M_{2},M_{3})$ are unaffected by the prior measurement $\hat{M_1}$, so that,
\begin{eqnarray}
NSIT_{(1)23}:P(M_{2},M_{3})&=&\sum_{M_{1}} P(M_{1},M_{2},M_{3})
\end{eqnarray}

 It is recently argued \cite{maroney,clemente} that along with the NSIT conditions the arrow-of-time(AoT) conditions are also required for LGIs and for macrorealism. The AoT condition can read as
\begin{eqnarray}
\label{aot1}
AoT_{i(j)}:P(M_{i})=\sum_{M_{j}} P(M_{i},M_{j}).
\end{eqnarray}
 which implies that the measured probability $P(M_{i})$ is unaffected by a future measurement $M_{j}$.  Similarly, $AoT_{12(3)}$ can be written as 
 
\begin{eqnarray}
AoT_{12(3)}:P(M_{1},M_{2})&=&\sum_{M_{3}} P(M_{1},M_{2},M_{3}).
\end{eqnarray}

Since no information can travel back in time, AoT conditions are naturally satisfied and irrelevant to the present discussion. Now, $NSIT_{1(2)3}$ is particularly interesting which can be written as 
 \begin{eqnarray}
NSIT_{1(2)3}:P(M_{1},M_{3})=\sum_{M_{2}} P(M_{1},M_{2},M_{3}).
\end{eqnarray}
This is actually the combination of $NSIT_{(2)3}$ and $AoT_{1(2)}$.

 Maroney and Timpson \cite{maroney} have shown that three-time NSIT conditions are necessary for SLGI. Note  here that a more general term `operational nondisturbance' is used in Ref.\cite{maroney} in place of NSIT. One can then write  
\begin{eqnarray}
\label{lg123}
NSIT_{(1)23}\wedge NSIT_{1(2)3} \Rightarrow SLGIs
\end{eqnarray}
The above implications is strictly unidirectional because the satisfaction of SLGI does not imply one or both the NSIT conditions are satisfied. Since WLGIs use the joint probabilities similar to SLGIs, the three-time NSIT conditions are also necessary for WLGIs too. 

Very recently, Clemente and Kofler \cite{clemente} have argued that  although three-time NSIT conditions( $NSIT_{1(2)3}$ and $NSIT_{(1)23}$)  are necessary and sufficient for SLGIs but \textit{not} for the macrorealism - a feature, which is in sharp contrast to the relationship between Bell inequality and local realism.  However, the conjunction of suitably chosen two-time and three-time NSIT conditions provides the necessary and sufficient condition for macrorealism. They argue that 
\begin{eqnarray}
NSIT_{(2)3} \wedge NSIT_{(1)23} \wedge NSIT_{1(2)3} \Leftrightarrow MR_{123}
\end{eqnarray}
The choice of two-time NSIT condition is  not unique. One may replace  $NSIT_{(2)3}$ by $NSIT_{(1)3}$. 

We now closely examine the connection between NSIT conditions, WLGIs and macrorealism for the sharp measurement  in the context of three-time LG measurement scenario considered in this paper. While no set of SLGIs can provide the necessary sufficient condition for macrorealism \cite{clemente16}, our study reveals that if all NSIT conditions are violated then the violation of at least one of the $24$ WLGIs  can be obtained for any state and for any value of $\tau$. In order to explore this, it is helpful to quantify the NSIT conditions. The violation of a NSIT condition occurs if the relevent prior measurement disturbs the subsequent mesurements\cite{maroney}. 

Let us consider the pair-wise marginal statistics of the experimental arrangement when all three measurements( $M_1$, $M_2$ and $M_3$) are performed. So that, one can write
\begin{eqnarray}
\label{triple1}
P_{(M_{1},M_{2},M_{3})}(M_{1}^{i},M_{2}^{j})&=&\sum_{k} P_{(M_{1},M_{2},M_{3})}(M_{1}^{i},M_{2}^{j},M_{3}^{k})\nonumber\\
\end{eqnarray}
\begin{eqnarray}
\label{triple2}
P_{(M_{1},M_{2},M_{3})}(M_{2}^{j},M_{3}^{k})&=&\sum_{i} P_{(M_{1},M_{2},M_{3})}(M_{1}^{i},M_{2}^{j},M_{3}^{k})\nonumber\\
\end{eqnarray}
\begin{eqnarray}
\label{triple3}
P_{(M_{1},M_{2},M_{3})}(M_{1}^{i},M_{3}^{k})&=&\sum_{j} P_{(M_{1},M_{2},M_{3})}(M_{1}^{i},M_{2}^{j},M_{3}^{k})\nonumber\\
\end{eqnarray}
where $i,j,k=\pm$.\\

Similarly, we consider single marginal statistics when two measurements are performed. We then have 
\begin{eqnarray}
\label{double1}
P_{(M_{1},M_{2})}(M_{2}^{j})&=&\sum_{i} P_{(M_{1},M_{2})}(M_{1}^{i},M_{2}^{j})
\end{eqnarray}
\begin{eqnarray}
\label{double2}
P_{(M_{1},M_{3})}(M_{3}^{k})&=&\sum_{i} P_{(M_{1},M_{3})}(M_{1}^{i},M_{3}^{k})
\end{eqnarray}
\begin{eqnarray}
\label{double3}
P_{(M_{2},M_{3})}(M_{3}^{k})&=&\sum_{j} P_{(M_{2},M_{3})}(M_{2}^{j},M_{3}^{k})
\end{eqnarray}

Now we define the following quantities
\begin{eqnarray}
\label{d11}
D_{1}(M_{2}^{j},M_{3}^{k})=P(M_{2}^{j},M_{3}^{k})-P_{(M_{1},M_{2},M_{3})}(M_{2}^{j},M_{3}^{k})
\end{eqnarray}
\begin{eqnarray}
\label{d22}
D_{2}(M_{1}^{i},M_{3}^{k})=P(M_{1}^{i},M_{3}^{k})-P_{(M_{1},M_{2},M_{3})}(M_{1}^{i},M_{3}^{k})
\end{eqnarray}
\begin{eqnarray}
\label{t21}
D_{1}(M_{2}^{j})=P(M_{2}^{j})-P_{(M_{1},M_{2})}(M_{2}^{j})
\end{eqnarray}
\begin{eqnarray}
\label{t31}
D_{1}(M_{3}^{k})=P(M_{3}^{k})-P_{(M_{1},M_{3})}(M_{3}^{k})
\end{eqnarray}
\begin{eqnarray}
\label{t32}
D_{2}(M_{3}^{k})=P(M_{3}^{k})-P_{(M_{2},M_{3})}(M_{3}^{k})
\end{eqnarray}

where $D_{1}(M_{2}^{j},M_{3}^{k})$ quantifies the amount of violation of three-time NSIT condition if the measurement $M_1$ at $t_1$ is performed. Similarly for $D_{2}(M_{1}^{i},M_{3}^{k})$. The quantity $D_{2}(M_{3}^{k})$ quantifies the amount of violation of two-time NSIT condition due to the measurement of $M_2$ at $t_2$ and similarly for $D_{1}(M_{3}^{k})$ and $D_{1}(M_{2}^{j})$.

For the pure state $|\psi(t_{1})\rangle$ given by Eq.(\ref{state}), the quantities given by Eqs.(\ref{d11}-\ref{t32}) can be derived as
\begin{eqnarray}
\label{ns1}
D_{2}(M_{1}^{+},M_{3}^{+})=-D_{2}(M_{1}^{+},M_{3}^{-})=-\frac{\cos ^2\theta  (\sin 2\tau)^2}{2}\nonumber\\ 
\end{eqnarray}
\begin{eqnarray}
\label{ns2}
D_{2}(M_{1}^{-},M_{3}^{+})=-D_{2}(M_{1}^{-},M_{3}^{-})=\frac{\sin^2\theta (\sin 2\tau)^2}{2}
\end{eqnarray}

\begin{eqnarray}
\label{ns3}
D_{1}(M_{2}^{+},M_{3}^{+})=-D_{1}(M_{2}^{-},M_{3}^{-})=\cos^3\tau \sin 2\theta \sin \tau \sin \phi\nonumber\\  
\end{eqnarray}

\begin{eqnarray}
\label{ns4}
D_{1}(M_{2}^{-},M_{3}^{+})=-D_{1}(M_{2}^{+},M_{3}^{-})=-\sin^3\tau  \sin 2\theta \cos \tau \sin \phi \nonumber\\ 
\end{eqnarray}

\begin{eqnarray}
\label{ns5}
D_{1}(M_{2}^{+})=-D_{1}(M_{2}^{-})=\frac{(\sin2\tau  \cos2\theta \sin \phi)}{2}
\end{eqnarray}
\begin{eqnarray}
\label{ns6}
D_{1}(M_{3}^{+})=-D_{1}(M_{3}^{-})=\frac{(\sin 2\theta \sin 4\tau \sin \phi)}{2}
\end{eqnarray}
\begin{eqnarray}
\label{ns7}
D_{2}(M_{3}^{+})&=&-D_{2}(M_{3}^{-})\nonumber\\ 
&=&\frac{(-2\sin^2\tau  \cos2\theta 
+\sin 2\theta \sin 4\tau \sin \phi)}{4}\nonumber\\
\end{eqnarray}
From the Eqs. (\ref{ns1}-\ref{ns7}) we can see that for $\phi\neq0$, $\theta\neq (\pi/4,0)$, and $\tau\neq\pi/4$, all the NSIT conditions are violated. We have found that in that situation, the violation of one of the $24$ WLGIs can be obtained. \\

Now, let us discuss the following interesting feature.  We can see from Eqs. (\ref{ns1}-\ref{ns7}) that for the values of $\theta=\pi/4$ and $\phi=0$ (i.e., $\rho(t_1)=|+\rangle\langle +|$)  we have $D_{2}(M_{1}^{+},M_{3}^{+})=D_{2}(M_{1}^{-},M_{3}^{-})=-D_{2}(M_{1}^{+},M_{3}^{-})=-D_{2}(M_{1}^{-},M_{3}^{+})=- (\sin^2 2\tau)/4$ but rest of the NSIT conditions are all satisfied. Similar feature can also be obtained for maximally mixed state ($I/2$). For the state $\rho(t_1)=|+\rangle\langle +|$ or $\rho(t_1)=I/2$, the quantum mechanical expressions for $24$ WLGIs take the form of one of them $\Delta_{w}^{Q}= \cos2\tau\sin^{2}\tau$, $-\sin^{2} 2\tau/2$, $-\cos^{2}\tau\cos2\tau$. It is then straightforward to see that the WLGIs are  not violated for those states at $\tau=\pi/4$ as shown in Figure $3$. But for those states $D_{2}(M_{1}^{i},M_{3}^{k})\neq 0$ at $\tau=\pi/4$. 

\begin{figure}[h]
{\includegraphics[width=8cm]{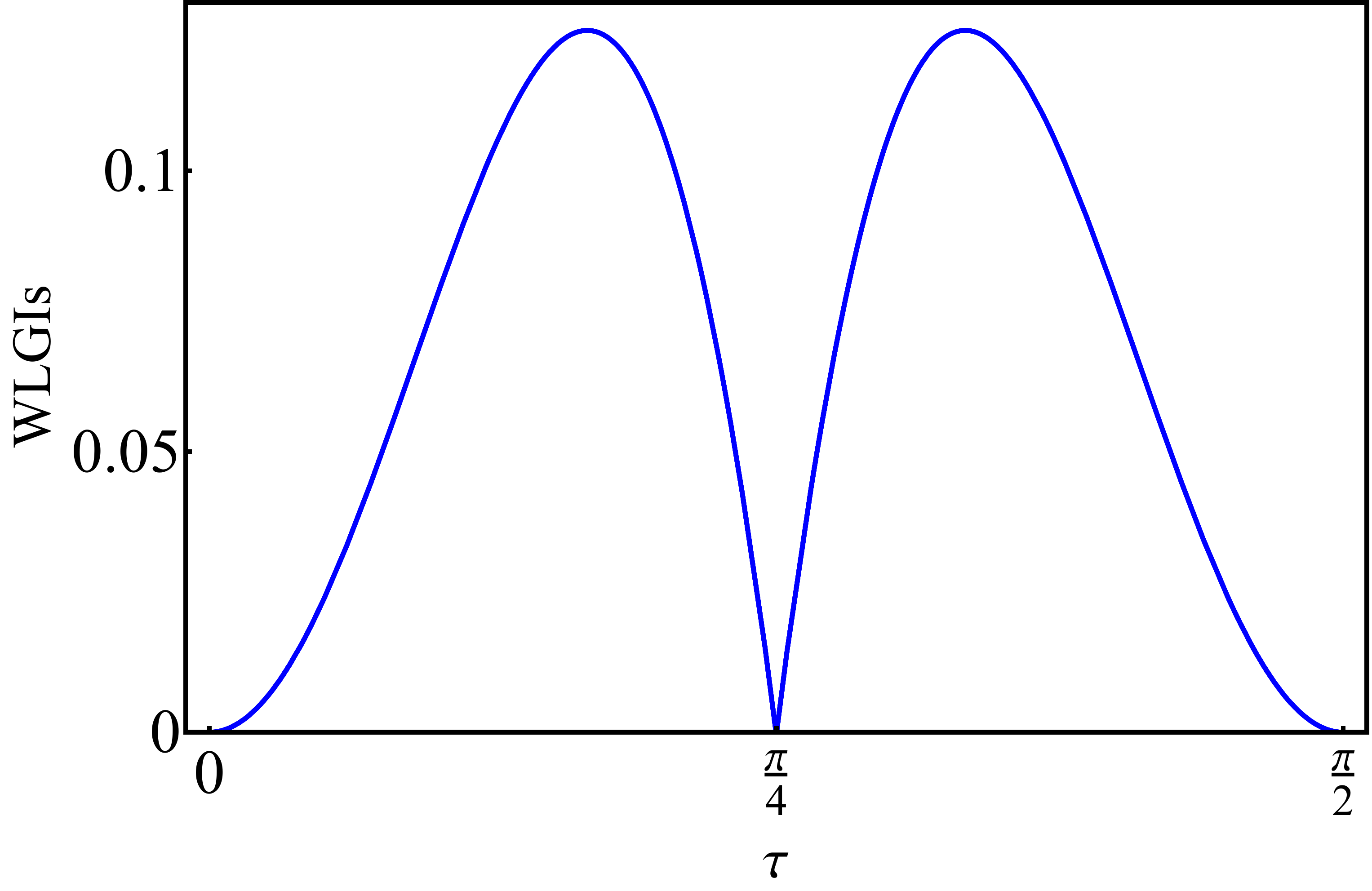}}
\caption{All $24$  WLGIs are plotted against $\tau$ for the state $\rho(t_0)=|+\rangle\langle +|$ or $\rho(t_0)=I/2.$ It is seen that at least one of the WLGIs are violated for all values of $\tau$ except at $\tau=\pi/4$.} 
\end{figure}
Based on the above study, we can thus claim that in the three-time LG scenario considered here, if  one of the three-time NSIT conditions is violated, then  one of the $24$ WLGIs can also be  violated except at $\tau=\pi/4$ and  for two instants, when $\rho(t_0)=I/2$  and $\rho(t_0)=|+\rangle\langle +| $. Note that, at $\tau=\pi/4$ the SLGIs (QM expressions of which are state independent) are always satisfied.\\

Next, we provide an analysis why no WLGI is violated for those specific instances. For this let us first take Eq.(\ref{wlgi}) as an example for providing a sketch of the argument. In order to obtain the violation of Eq.(\ref{wlgi}), the disturbance caused by the prior measurement to the future measurement plays an important role.
The $\Delta_{w}^{LG}$ given by Eq.(\ref{wlgi}) can be written for three measurement scenario as $(\Delta_{w}^{LG})_{M_{1}M_{2}M_{3}}=\sum_{M_{1}=\pm 1} P(M_{1},M_{2}^{+},M_{3}^{-})-\sum_{M_{3}=\pm 1} P(M_{1}^{+},M_{2}^{+},M_{3})-\sum_{M_{1}=\pm 1} P(M_{1}^{-},M_{2},M_{3}^{-})=-P(M_{1}^{+},M_{2}^{+},M_{3}^{+})-P(M_{1}^{-},M_{2}^{-},M_{3}^{-})$.  Clearly, if $\Delta_{w}^{LG}=(\Delta_{w}^{LG})_{M_{1}M_{2}M_{3}}$,  WLGI will not be violated.  Using above relations Eqs.(\ref{d11}) and (\ref{d22}), we can write $\Delta_{w}^{LG}-(\Delta_{w}^{LG})_{M_{1}M_{2}M_{3}}=D_{1}(M_{2}^{+},M_{3}^{-})-D_{3}(M_{1}^{+},M_{2}^{+})-D_{2}(M_{1}^{-},M_{3}^{-})$. 

Since AoT conditions are always satisfied, $D_{3}(M_{1}^{+},M_{2}^{+})=0$. By noting $\Delta_{w}^{LG}\leq 0$ in Eq.(\ref{wlgi}), we can write
	 \begin{eqnarray}
	&&D_{1}(M_{2}^{+},M_{3}^{-})-D_{2}(M_{1}^{-},M_{3}^{-})\\
	\nonumber
	&-&P(M_{1}^{+},M_{2}^{+},M_{3}^{+})-P(M_{1}^{-},M_{2}^{-},M_{3}^{-})\leq 0
	\end{eqnarray}
The WLGI given by Eq.(\ref{wlgi}) is violated if the condition
	 \begin{eqnarray}
	\label{dh4v}
	D_{1}(M_{2}^{+},M_{3}^{-})-D_{2}(M_{1}^{-},M_{3}^{-})>P(M_{1}^{+},M_{2}^{+},M_{3}^{+})\\
	\nonumber
	+P(M_{1}^{-},M_{2}^{-},M_{3}^{-})
	\end{eqnarray}
is satisfied.  Similar $23$ more inequalities (corresponding to the other $23$ WLGIs) can be derived in such a manner. If we write them in the compact notations, 
\begin{eqnarray}
\label{dh4s}
	D_{1}(M_{2}^{j},M_{3}^{k})-D_{2}(M_{1}^{i},M_{3}^{k}) > P(M_{1}^{i},M_{2}^{-j},M_{3}^{k})\\
	\nonumber
	+P(M_{1}^{-i},M_{2}^{j},M_{3}^{-k})
	\end{eqnarray}
	\begin{eqnarray}
	\label{d2}
	D_{2}(M_{1}^{i},M_{3}^{k})-D_{1}(M_{2}^{j},M_{3}^{k}) >P(M_{1}^{-i},M_{2}^{j},M_{3}^{k})\\
	\nonumber
	+ P(M_{1}^{i},M_{2}^{-j},M_{3}^{-k})
	\end{eqnarray}
	\begin{eqnarray}
	\label{d3}
	-D_{2}(M_{1}^{i},M_{3}^{k})-D_{1}(M_{2}^{j},M_{3}^{-k}) > P(M_{1}^{i},M_{2}^{-j},M_{3}^{k})\\ 
	\nonumber
	+P(M_{1}^{-i},M_{2}^{j},M_{3}^{-k})
	\end{eqnarray}
	 
The quantity  $D_{1}(M_{2}^{+},M_{3}^{-})=P(M_{2}^{+},M_{3}^{-})-P(M_{1}^{+},M_{2}^{+},M_{3}^{-})-P(M_{1}^{-},M_{2}^{+},M_{3}^{-})$ quantifies the amount of violation of $NSIT_{(1)23}$ while obtaining the outcome $M_{2}=+1$ and $M_{3}=-1$ and similarly for others.  If the measurement at $t_1$ does not produce any disturbance to the measurements at $t_2$ and $t_3$, then $D_{1}(M_{2}^{+},M_{3}^{-})=0$. Similar argument holds good for $D_{2}(M_{1}^{-},M_{3}^{-})$. Note that, for the violation of WLGI given by Eq.(\ref{wlgi}), at least one  of the three-time NSIT conditions has to be violated. Then the three-time NSIT conditions are necessary for WLGIs. However, they are not sufficient. It can be seen from ineq.(\ref{dh4v}) that the mere violation of three-time NSIT condition is not enough for the violation of WLGI. An interplay between different NSIT conditions and a threshold value plays the key role.

If a set of WLGIs are violated then corresponding set of inequalities ineqs.(\ref{dh4s}-\ref{d3}) need to be satisfied. We have already shown that in  three-time LG scenario,  for two specific states of $\rho(t_0)=I/2$ and $\rho(t_0)=|+\rangle\langle +|$ at $\tau = \pi/4$, none of the  ineqs.(\ref{dh4s}-\ref{d3}) is satisfied (meaning that no WLGI is violated). For $\tau=\pi/4$, the measured observables at time $t_1$, $t_2$ and $t_3$ become $\sigma_{z}$, $\sigma_{y}$ and $\sigma_{z}$ respectively. It is then straightforward to understand that for the states $\rho(t_0)=I/2$ and $\rho(t_0)=|+\rangle\langle +|$, any three-time joint probability $P(M_{1}^\pm, M_{2}^\pm , M_{3}^\pm )$ is equal to $1/8$, leading every right hand side of ineqs.(\ref{dh4s}-\ref{d3}) to $1/4$. The values of $D_{1}(M_{2}^{\pm},M_{3}^{\pm})$ and $D_{2}(M_{1}^{\pm},M_{3}^{\pm})$ ranges from $-1/4$ to $1/4$. For maximally mixed state $\rho(t_0)=I/2$, the measurement at $t_1$ can cause \emph{no} disturbance to the subsequent measurements implying $D_{2}(M_{1}^{\pm},M_{3}^{\pm})=0$ and for the state $\rho(t_0)=|+\rangle\langle +|$, after the measurement at $t_1$ the reduced state becomes a maximally mixed states, then $D_{1}(M_{2}^{+},M_{3}^{-})=0$. This then explains why for those particular state at $\tau=\pi/4$ no  violation of any of the WLGIs is obtained. 

We have also analyzed whether for a more general observables and evolutions one can get violation of one of the WLGIs for those aforementioned instances. We found that for a different Hamiltonian the violation of WLGIs can be obtained for $\rho(t_0)=|+\rangle\langle +|$ or $\rho(t_0)=I/2$ at $\tau=\pi/4$. But, in such a case there can be other states and different values of $\tau$ for which no violation of any of the WLGIs will be obtained.	A simple example can be helpful. Let us consider the observable $M_1=\sigma_z$ and evolution Hamiltonian $H=\omega(\cos\alpha\sin\beta\sigma_x+\cos\alpha\cos\beta\sigma_y+\sin\alpha\sigma_z)$. By arbitrarily choosing $\alpha=\beta=\pi/4 $,  we found the violation of one of the WLGIs at $\tau=\pi/4$ for the above mentioned states. But no violation of any of the WLGIs can be obtained for a significant ranges of values around $\tau=\pi/3$ (Figure. 4) for the state $\rho(t_0)=|0\rangle\langle 0|$(or $\rho(t_0)=|1\rangle\langle 1|$). Note that, for those states $D_{2}(M_{1}^{i},M_{3}^{k})\neq 0$ at $\tau=\pi/3$. Same argument holds good for maximally mixed state. Then for more general time evolution, violation of any of the WLGIs does not occur for a considerably larger range of $\tau$ compared to only at $\tau=\pi/4$ in the earlier choice of Hamiltonian. 
\begin{figure}[h]
{\includegraphics[width=8cm]{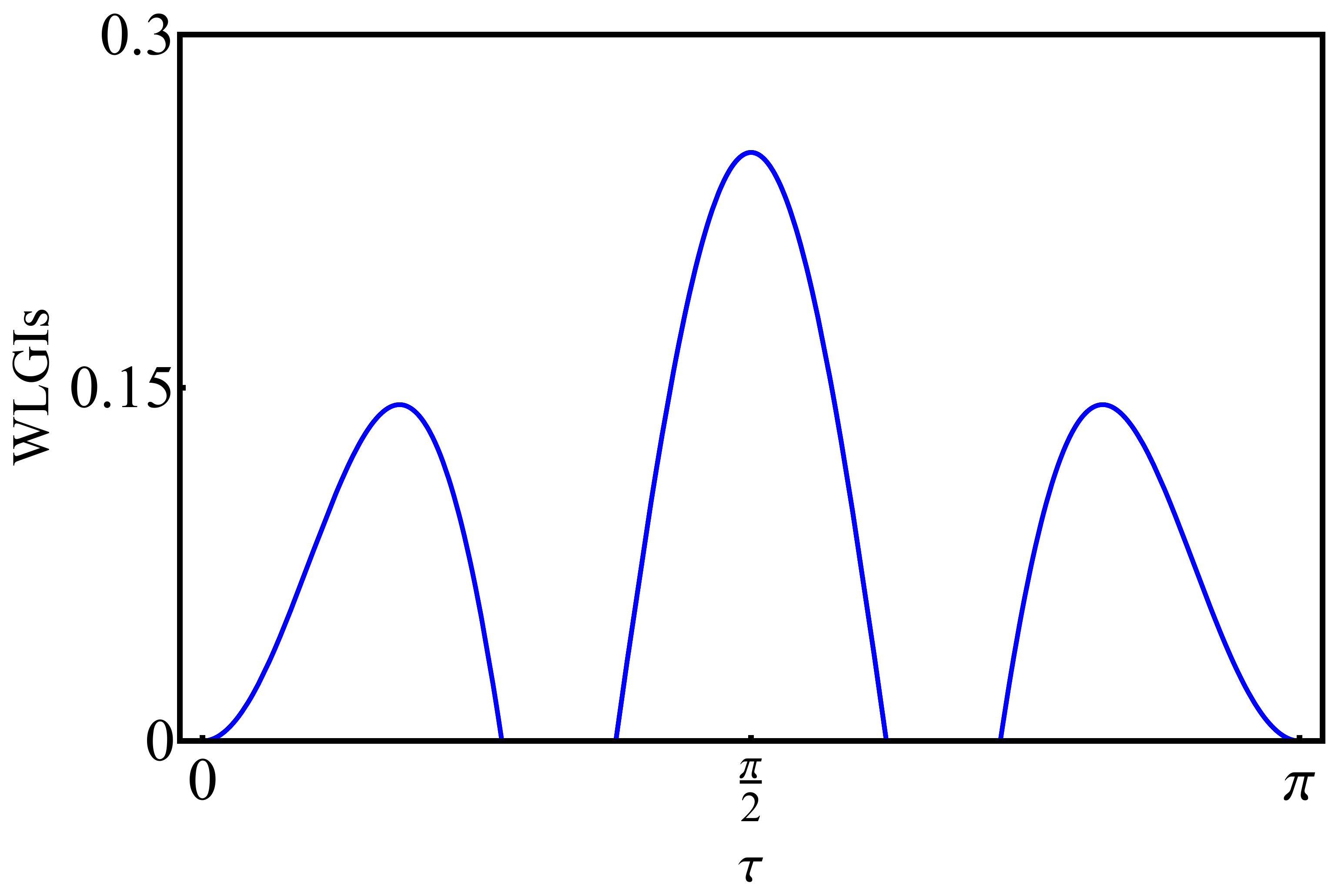}}
\caption{All $24$  WLGIs are plotted against $\tau$ for the state $\rho(t_0)=|0\rangle\langle 0|$. It is seen that no WLGI is violated for a considerable range around $\tau=\pi/3$. } 
\end{figure}

It is recently proved by us that WLGIs are stronger than SLGIs \cite{kumari}, i.e., WLGIs provide a better test of macrorealism than the SLGIs. But we have seen here that they also do not provide the necessary and sufficient condition for macrorealism in the three-time LG scenario. However, a suitable conjunction of NSIT conditions  fully captures \cite{clemente} the notion of macrorealism for any arbitrary state and measurement scenario. Then the violation of one of the NSIT conditions provide the violation of macrorealism. In contrast, the violation of a LGI requires an interplay between different three-time NSIT conditions and a threshold value, as can be seen from ineqs.(\ref{dh4s}-\ref{d3}). We provided an explanation with the help of the notion of the disturbance why NSIT condition is better candidate than LGIs for testing macrorealism. 

\section{Joint Measurability of POVMs and Macrorealism}
Fine's theorem says that the following statements are equivalent.\\

1) There exists a global joint probability distribution for all outcomes whose marginals are the experimentally observed probabilities. 2) There exists a local realistic model for all probabilities. 3) All Bell inequalities are satisfied.

Note that the SLGI is often considered to be the temporal analogue of Bell's inequalities. But it is recently shown \cite{clemente16} that \emph{no} set of SLGI can provide necessary and sufficient condition for macrorealism, i.e., second and third statements of Fine's theorem\cite{fine} are inequivalent for three-time LG scenario for testing macrorealism.  The purpose of this section is to examine the equivalence between first and third statements. In particular, the possible connection between the joint measurability and the violation of LGIs is probed.

A couple of brief attempts have been made along this direction \cite{emary,kartik}. For the case of sharp measurements, the non-joint measurability of two observables of SLGI given by ineq.(\ref{lgi}) implies the notion of non-commutativity. The non-commutativity of sharp observables $\hat{M}$ at different times satisfy the commutation relation $[\hat{M_i}, \hat{M_j}]=2 i \hat{\sigma}.(m_i\times m_j)$, where the vectors $m_i$ are all lie in the $x-z$ plane with equal angles($\tau$) between them. Then $[\hat{M_1}, \hat{M_2}]=[\hat{M_2}, \hat{M_3}]=2 i \hat{\sigma_y} \sin\tau$ and $[\hat{M_1}, \hat{M_3}]=2 i \hat{\sigma_y} \sin 2\tau$. Emary \textit{et al}.\cite{emary} have argued that the values of $\tau$, where the commutators simultaneously vanish are the values where violation of  SLGI disappears. For unsharp measurement in case of trine-spin POVMs it has been shown \cite{kartik}  that triple-wise joint measurability condition  is related to the  violation of SLGI \textit{type} inequality. This example \cite{kartik} did not consider the time correlations and hence \textit{not} directly related to the spirit of the notion of macrorealism. Here, we found that even when POVMs are compatible the violation of a WLGI can be obtained. In order to demonstrating this, let us consider the joint  measurement conditions for two different POVMs, $M^{\pm} (x,\vec{m})$ and $M^{\pm} (y,\vec{n})$. The general condition of pair-wise joint measurability \cite{yu} is the following;
\begin{equation}
\label{jme}
(1-F_{x}^{2}-F_{y}^{2})(1-\frac{x^2}{F_{x}^{2}}-\frac{y^2}{F_{y}^{2}})\leq(\vec{m}.\vec{n}- x y)^2
\end{equation}
where $F_{x}$ and $F_{y}$ are given by\\
\begin{equation}
F_{x}=\frac{\sqrt{(1-x)^2-m^2}+\sqrt{(1+x)^2-m^2}}{2};
\end{equation}
\begin{equation}
F_{y}=\frac{\sqrt{(1-y)^2-n^2}+\sqrt{(1+y)^2-n^2}}{2};
\end{equation}
For $x=0$ and $y=0$ we obtain the  well-known joint measurability condition \cite{busch} for the spin-POVMs is given by
\begin{equation}
\label{jms}
||\vec{m}+\vec{n}||+||\vec{m}-\vec{n}||\leq2
\end{equation}
Using Eq.(\ref{jms}), the pair-wise joint measurability condition for our aforementioned POVMs $M_1^{\pm}$ and $M_2^{\pm}$ (and for $M_2^{\pm}$ and $M_3^{\pm}$) can be obtained as
\begin{equation}
\label{jm1}
\eta \leq (\cos\tau + \sin\tau)^{-1}
\end{equation}
Similarly, pair-wise joint measurability condition for $M_1^{\pm}$ and $M_3^{\pm}$ is given by 
\begin{equation}
\label{jm2}
\eta \leq (\cos 2\tau + \sin 2\tau)^{-1}
\end{equation}
 The minimum value of RHS in Eq.(\ref{jm1}) and Eq.(\ref{jm2}) can be $0.707$ at two different times. Then, POVMs $M_{i}^{\pm}(x,\vec{m_{i}})$ where $i=1,2,3$ are pair-wise jointly measurable when $\eta\leq 0.707$ is satisfied. It can be seen from Eq.(\ref{sl1}) that  the violation of SLGI occurs only when $\eta>0.81$. So there is gap between $0.707-0.81$ where the violation of SLGI does \textit{not} occur. Similar inference can be made for the violation of ELGI. However, the WLGI is violated at $\eta>0.69$. Thus, WLGIs can be violated even when the POVMs are pair-wise jointly measurable. 

Let us now examine the triple-wise joint measurability condition of spin-POVMs. The triple-wise joint measurability condition for spin- POVMs  $M^{\pm1}(0,\vec{m_1}), M^{\pm2}(0,\vec{m_2})$ and $ M^{\pm3}(0,\vec{m_3})$ can be obtained from the following condition \cite{son},
\begin{eqnarray}
\label{jmw4}
&&|\vec{m_1}+\vec{m_2}+\vec{m_3}|+|\vec{m_1}+\vec{m_2}-\vec{m_3}|\\
\nonumber
&&+|\vec{m_1}-\vec{m_2}-\vec{m_3}|+|\vec{m_1}-\vec{m_2}+\vec{m_3}|\leq4
\end{eqnarray}
For the spin-POVMs used in the context of LG scenario, from Eq.(\ref{jmw4}), the triple-wise joint measurability condition to be $\eta\leq 0.54$. This indicates that the violation of LGIs may not be obtained, when the spin-POVMs are triple-wise incompatible.
 
We now consider  biased-POVMs when $x=\eta-1$. Using Eq.(\ref{jme}), the pair-wise joint-measurability condition for $M_1^{\pm}$ and $M_2^{\pm}$ (and for $M_2^{\pm}$ and $M_3^{\pm}$) can be obtained as
\begin{equation}
\label{jm3}
\eta\leq (1+ \cos\tau)^{-1}
\end{equation}
and for $M_2^{\pm}$ and $M_3^{\pm}$ joint measurability condition is 
\begin{equation}
\label{jm4}
\eta\leq (1+ \cos2\tau)^{-1}
\end{equation}
Note that there is a discontinuity in ineqs.(\ref{jm3}-\ref{jm4}). For $\tau=0$, we have $\eta\leq 1/2$. But in that case the two POVMs are same. For detailed discussion of this issue, we refer Ref.\cite{yu,teiko,p.busch}.

It is already shown in the earlier Sections that for biased POVMs, the SLGI, WLGI and ELGI given by ineq.(\ref{sl2}), (\ref{wlgi}) and (\ref{MRe}) respectively are violated for \textit{any nonzero value of} $\eta$. From Eq.(\ref{jm3}) and Eq.(\ref{jm4}) we see that two POVMs are pair-wise jointly measurable for $\eta\leq 0.589$. Therefore, there is \emph{no} connection between pair-wise joint measurability of biased POVMs and violation of LGIs. The triple-wise joint measurability of biased POVMs is not known. Since all formulation of LGIs are violated for any non-zero value of $\eta$, no connection may be found between triple-wise joint measurability and violation of LGIs.

\section{Summary and discussions}
In this paper, we first provided a detailed study of the violations of various formulations of LGIs, viz., standard LGI (SLGI), Wigner form of LGIs (WLGIs) and entropic LGI (ELGI) for the case of unsharp measurements. While for the case of spin-POVMs the violations of WLGIs are more robust than that of  SLGIs and ELGIs, our study reveals that  for the case of biased POVMs the violations of SLGIs and ELGIs provide the same robustness as WLGIs. Importantly, the violation of all formulations of LGIs for biased POVMs can be achieved for  \textit{any non-zero value} of unsharpness parameter($\eta$). We thus demonstrated that if LGIs is taken to be an indicator of classicality of a macroscopic system then it does \textit{not} emerge for any arbitrary unsharp measurement. As regards the realizability of biased POVM, we remark that according to Naimark's theorem \cite{peres} any POVM can be realized by extending the Hilbert space to a larger space and then by performing projective measurements. 

We have also  examined the connection between the WLGIs, NSIT conditions and macrorealism. By invoking the  notion of disturbance, we explained why NSIT conditions provide better test of macrorealism than LGIs which is in accordance with a recent claim \cite{clemente}. Note that for two-party, two-measurement and two-outcome Bell Scenario, the CHSH inequalities provide the necessary and sufficient condition for local realism \cite{fine}. Although SLGIs seem to be the temporal analogue of CHSH inequalities but no set of SLGIs can provide necessary and sufficient condition for macrorealism \cite{clemente16}. However, a suitable  conjunction of NSIT conditions provide the same.  For the usual two-qubit Bell scenario the only relevant inequality is the CHSH one. We have recently provided a generic proof to show that WLGIs are inequivalent and stronger than SLGIs \cite{kumari}. It is then an interesting question whether WLGIs provide the necessary and sufficient condition for macrorealism. Pertaining to the three-time LG scenario, we have found that if at least one of the three-time NSIT conditions is violated then one of the $24$ WLGIs is  violated for almost all states except for two specific instances.  Then WLGIs also do not provide necessary and sufficient condition for macrorealism. Conclusion remains same for the more general observables and intermediate evolutions. Further, we provided an interesting reasoning why no WLGI is violated for those instances. This is argued by invoking the notion of the amount of the violation of various NSIT conditions. We showed that although the three-time NSIT conditions are necessary for LGIs but mere violation of them do not warrant the violations of LGIs.  This is due to the fact that for the violation of a particular WLGI (or SLGI), an interplay between the violation of NSIT conditions and a threshold value involving three-time joint probabilities plays an important role.

In his celebrated work, Fine\cite{fine} demonstrated the connection between local joint measurability and two-party, two-measurement and two-outcome CHSH inequalities. We have made a detailed study here to examine whether a similar connection can be established between pair-wise or triple-wise joint measurability and the violation of LGIs. Our study reveals that there is no such generic connection. 

In a recent study  \cite{chaves} it is argued that the entropic inequalities can provide the necessary and sufficient condition for non-contextuality and locality. It would then be interesting to examine if ELGIs  provide the same for macrorealism. In view of our study, it would also be instructive to formulate new set of inequalities to examine whether that new set along with the existing set of LGIs provide necessary and sufficient condition for macrorealism.  A study along this direction is very recently initiated \cite{hall} by using a quasi-probability approach. A comparison is also made \cite{hal2}between the LGIs and NSIT conditions by introducing the notion of weak and strong macrorealism where a somewhat different formulation of LGIs is invoked and it is shown that such a form of LGIs provide the necessary and sufficient condition for weak macrorealism. Studies along this line could be an interesting avenue of research that will be carried out in future.    
\section*{Acknowledgments}
SK acknowledges the Research Assistantship of NIT Patna. AKP acknowledges the support from Ramanujan Fellowship research grant (SB/S2/RJN-083/2014). We are thankful to Johannes Kofler for useful discussions. We would like to thank an anonymous referee for her/his constructive criticisms and comments for improving the quality of the manuscript.

\end{document}